\documentclass[12pt]{iopart}

\usepackage{iopams}
\usepackage{ctex}
\usepackage{graphicx}
\begin{document}

\title[\JPCM]{Quantum phase transitions and string orders in the spin-1/2 Heisenberg-Ising alternating chain with Dzyaloshinskii-Moriya interaction}

\author{Guang-Hua Liu$^{1}$, Wen-Long You$^{2}$, Wei Li$^{3,\dag}$, and Gang Su$^{4}$}

\address{1 Department of Physics, Tianjin Polytechnic University, Tianjin 300387, P. R. China\\
2 College of Physics, Optoelectronics and Energy, Soochow University, Suzhou, Jiangsu 215006, P. R. China\\
3 Department of Physics, Beihang University, Beijing 100191, P. R. China\\
4 Theoretical Condensed Matter Physics and Computational Materials Physics Laboratory, College of Physical Sciences, University of Chinese Academy of Sciences, P. O. Box 4588, Beijing 100049, P. R. China}
\ead{$^{\dag}$ W.Li@physik.lmu.de}
\begin{abstract}
Quantum phase transitions (QPTs) and the ground-state phase diagram of the spin-1/2 Heisenberg-Ising alternating chain (HIAC) with uniform Dzyaloshinskii-Moriya (DM) interaction are investigated by a matrix-product-state (MPS) method. By calculating the odd- and even-string order parameters, we recognize two kinds of Haldane phases, i.e., the odd- and even-Haldane phases. Furthermore, doubly degenerate entanglement spectra on odd and even bonds are observed in odd- and even-Haldane phases, respectively. A rich phase diagram including four different phases, i.e., an antiferromagnetic (AF), AF stripe, odd- and even-Haldane phases, is obtained. These phases are found to be separated by continuous QPTs: the topological QPT between the odd- and even-Haldane phases is verified to be continuous and corresponds to conformal field theory with central charge $c$=1; while the rest phase transitions in the phase diagram are found to be $c$=1/2. We also revisit, with our MPS method, the exactly solvable case of HIAC model with DM interactions only on odd bonds, and find that the even-Haldane phase disappears, but the other three phases, i.e., the AF, AF stripe, and odd-Haldane phases, still remain in the phase diagram. We exhibit the evolution of the even-Haldane phase by tuning the DM interactions on the even bonds gradually.
\end{abstract}

\vspace{2pc} \noindent{\it Keywords}: iTEBD, QPTs, string order \\
\submitto{\JPCM}
\maketitle

\section{Introduction}
\label{sec1}

It is well known that the ground-state properties are intrinsically different between the uniform spin-1/2 and spin-1 Heisenberg antiferromagnetic chains (HAFCs). Specifically, the ground state of the former (spin-1/2) model is critical and has gapless low-energy excitations, \textit{i.e.}, spinons with fractional spin \cite{Mourigal-2013}; while for the latter spin-1 HAFC, there exists a finite energy gap, \textit{i.e.}, the Haldane gap, above the ground state, and the elementary excitations are magnons with nonzero mass\cite{White-1993}. This intriguing feature of the spin-1 HAFC was first conjectured by Haldane\cite{Haldane}, and then theoretically interpretated in terms of the valence bond solid state picture \cite{Affleck}. In addition, a nonlocal string order parameter $O_s$ has been introduced to characterize the ``dilute" antiferromagnetic (AF) order in the Haldane phase \cite{Nijs}. Further study shows that the topological long-range order in the $SO(2n+1)$ symmetric matrix product state (MPS) can be fully identified and characterized by a set of nonlocal string order parameters \cite{Tu}. The $SO(2n+1)$ symmetric MPS contains diluted AF orders in $n$ different channels and a hidden ($Z_2$$\times$$Z_2$)$^{n}$ symmetry breaking, which opens an excitation gap. The existence of a nonlocal order parameter and a dilute AF order actually are physically observables: the experimental observation of spin-1/2 edge states in spin-1 chains evidences the valence bond picture in the Haldane phase \cite{Hagiwara,Glarum}, and it is worthy noticing that people recently succeed to observe directly the nonzero string order in cold atom systems \cite{Endres}.

According to Landau's phase transition theory \cite{Landau-1937, Landau-1958}, different phases and their phase transitions can be characterized by local order parameters reflecting different symmetry breakings. However, the Haldane phase, which has no explicit symmetry breaking, is separated from a "trivial" phase, e.g., large-$D$ phase ($D$ represents single-ion anisotropy), by a phase transition \cite{Wen-2009,Pollmann-2010,Pollmann-2012}, as long as symmetries [SO(3), bond-centered inversion, etc] are intact in the Hamiltonian space (path). So to speak, the Haldane phase has a symmetry-protected topological (SPT) order \cite{Wen-2012}, and the string order parameter $O_s$ can thus be regarded as a quantification of the topological order of Haldane phase. Moreover, a nonlocal unitary transformation was constructed to uncover explicitly the hidden $Z_2$$\times$$Z_2$ symmetry breaking in the Haldane phase \cite{Tasaki, Kennedy,Oshikawa,Totsuka}.

On the other hand, SPT phases have also been reported in spin-1/2 multi-period chains. A spin-1/2 ferromagnetic-antiferromagnetic alternating chain has been shown to have nonzero string order, which changes continuously through the Haldane and dimer phases and show a crossover behavior between them \cite{Hida}. In addition, Kohmoto and Tasaki indicated that the spin-1/2 ferromagnetic-antiferromagnetic alternating chain fully breaks the hidden $Z_2$$\times$$Z_2$ symmetry not only in the strongly coupled (Haldane) phase but also in the decoupled (dimer) phase, supporting that the spin-1/2 dimer state belongs to a Haldane-like phase. Recently, Wang \textit{et al} \cite{Wang} associates the phase transition between an odd- and even-dimer states as topological quantum phase transition (TQPT) between two kinds of Haldane phases (even- and odd-Haldane), which have different string order parameters.

Among others, the Heisenberg-Ising alternating chain (HIAC), originally proposed by Lieb \textit{et al} \cite{Lieb} and re-examined subsequently by Yao \textit{et al} \cite{Yao}, constitutes an interesting spin-1/2 chain model. The Hamiltonian of the HIAC is given by
\begin{equation}
\hat{H} = \sum^{N/2}_{i} [J_{H} {\hat{\bold{S}}}_{2i-1} \cdot \hat{\bold{S}}_{2i} + J_{I} S^{z}_{2i} S^{z}_{2i+1}],
\label{Hamiltonian1}
\end{equation}
where $\hat{\bold{S}}$ denotes a spin-1/2 operator ($S^z$ is the $z$-component), and $N$ (an even number) is the total number of
spins. $J_{H}$ and $J_{I}$ ($J_{H}$, $J_{I}>0$)
are the Heisenberg and Ising couplings on the odd and even bonds, respectively. This model can be solved exactly, and a phase transition from a quantum paramagnetic phase to the AF phase was found to occur at $J_{I}/J_{H}=2$. When the Dzyaloshinskii-Moriya (DM) interactions are switched on odd bonds of the HIAC \cite{Strecka}, the model is still exactly solvable, it turns out that the ground-state energy exhibits an interesting non-analytic behavior accompanied by a gapless excitation spectrum along the line $J_{I}$=$2\sqrt{D^{2} + J^{2}_{H}}$, where $D$ denotes the strength of DM interaction [see Eq. (\ref{Hamiltonian2}) below]. This critical line was found to separate a quantum paramagnetic phase ($J_{I} < 2\sqrt{D^{2} + J^{2}_{H}}$) and an AF one. However, in both cases, the property of the SPT order has not been discussed in the paramagnetic phase.

In this paper, we study an extended HIAC model, which has an uniform DM interactions on both even- and odd-bonds [Eq.~(\ref{Hamiltonian2}) below], which is thus no longer exactly soluble. We here adopt a MPS based numerical method, i.e., the infinite time-evolving block decimation (iTEBD) algorithm\cite{Vidal03, Vidal}, to determine the ground state with high accuracy. We obtain a rich ground-state phase diagram which includes an AF phase, an antiferromagnetic stripe phase (AFSP), and two Haldane phases. In particular, the two Haldane phases, namely, the odd- and even-Haldane phases, can be characterized (and distinguished) by two different nonlocal string order parameters and doubly degenerate entanglement spectra on the odd and even bonds, respectively. Moreover, a continuous TQPT between the odd- and even-Haldane phases has been found.

The rest of this paper are arranged as follows. In section \ref{sec2}, the model Hamiltonian and the exact phase diagram is present. In section \ref{section3}, the numerical results are shown and analyzed in details. Lastly, we devote section \ref{sec4} to a discussion and summary.

\section{Model Hamiltonian and the phase diagram}
\label{sec2}

The spin-1/2 HIAC with an uniform DM interaction is described by
\begin{eqnarray}
 \hat{H} &=& \sum^{N/2}_{i} [J_{H} \hat{\bold{S}}_{2i-1} \cdot \hat{\bold{S}}_{2i}+ D (S^{x}_{2i-1} S^{y}_{2i}-S^{y}_{2i-1} S^{x}_{2i}) \nonumber \\
         &+& J_{I} S^{z}_{2i} S^{z}_{2i+1}+ D (S^{x}_{2i} S^{y}_{2i+1}-S^{y}_{2i} S^{x}_{2i+1})].
\label{Hamiltonian2}
\end{eqnarray}
Here $\vec{D}$ is called the DM vector, with $\vec{D}$=$D\hat{e}_z$ adopted in this paper. The DM interaction is an antisymmetric spin-spin coupling, due to the spin-orbit coupling effect.
Recently such DM-like spin-orbit interaction in the single-crystal yttrium iron garnet was experimentally verified and found amenable to manipulations \cite{Zhang14}. {It is worth noting the DM interaction can be eliminated by a canonical spin rotation \cite{Strecka,You11}, and then Eq. (\ref{Hamiltonian2}) can be mapped into a period-two XXZ model consequently. In contrast to Eq. (\ref{Hamiltonian1}), the transformed XXZ model is no longer exactly soluble.}

We employ the iTEBD method to investigate the ground-state properties of model Eq.~(\ref{Hamiltonian2}). The iTEBD algorithm is a numerical approach based on the MPS ansatz, which accurately describes quantum many-body wavefunctions in one dimension, given sufficiently large bond dimensions $\chi$. During iTEBD optimizations, one successively applies imaginary-time evolution gates $\exp{(-\tau h)}$ on a trial MPS state, until the latter converges to the variational ground state. $h$ is the local Hamiltonian (here only two-site coupling term on an even or odd bond), and $\tau$ is the Trotter step length. In practice, we start from $\tau=10^{-1}$ and gradually reduce it to smaller values (eventually down to $\tau=10^{-8}$), with a total evolution steps of number $O(10^{4\sim5})$, given the variational MPS wavefunction is well converged in the course of imaginary-time evolutions. Regarding the bond dimension $\chi$ of MPS used in practice, we found for this specific model, $\chi \sim 30$ can already provide rather accurate and converged results (for various gapped phases of the model). Remarkably, distinct from other finite-size algorithms like quantum Monte Carlo or finite-size density matrix renormalization group method, iTEBD exploits the translation invariance of the one-dimenional chain and make the thermodynamic limit directly accessible.

\begin{figure}
\begin{center}
\includegraphics[width=0.7\linewidth]{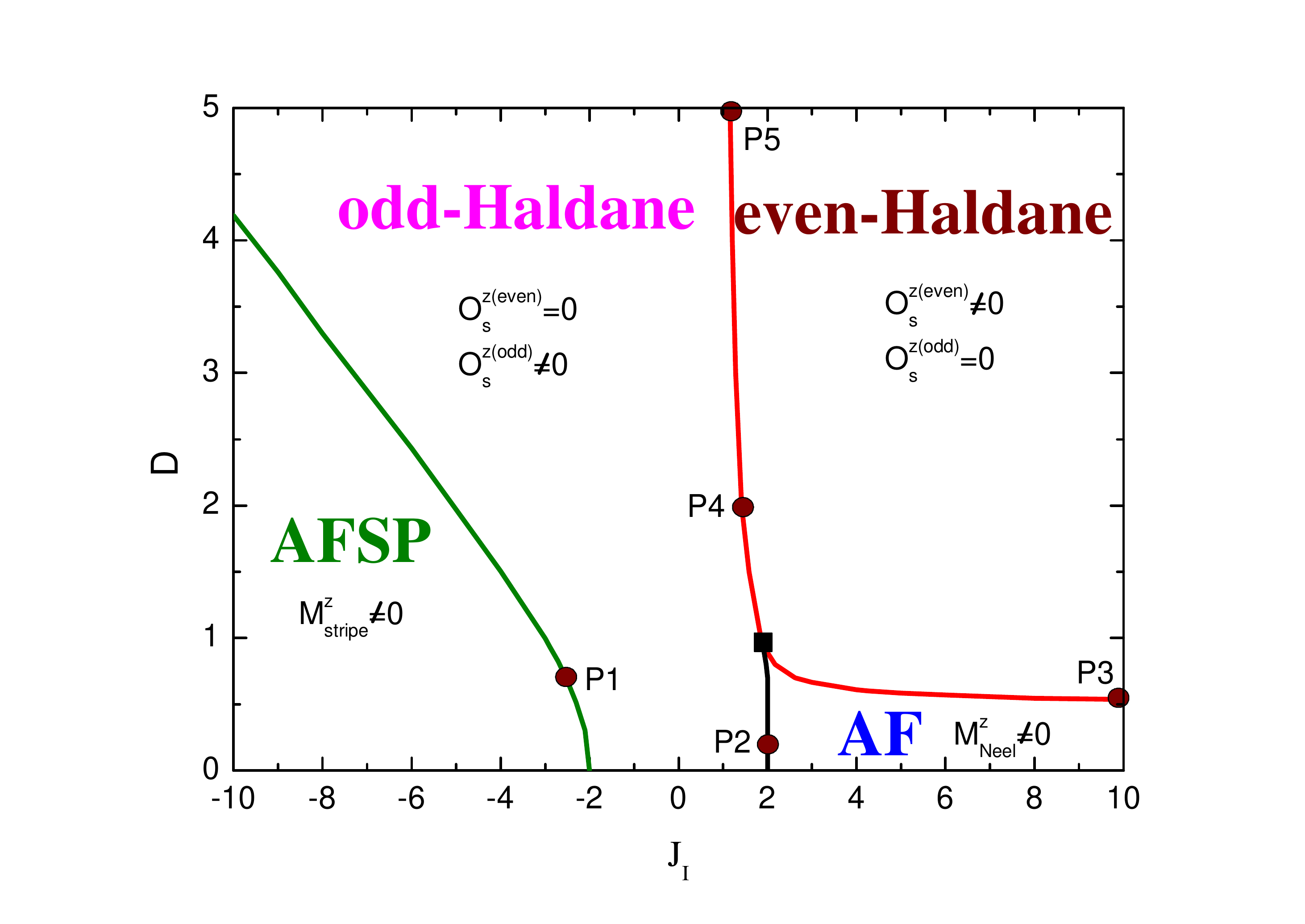}
\end{center}
\caption{(Color online) Magnetic phase diagram of the spin-1/2 HIAC with uniform DM interactions [Eq. (\ref{Hamiltonian2})]. It includes four different phases: an AFSP, an AF phase, and two kinds of Haldane (odd- and even-Haldane) phases. The filled square $\blacksquare$ denotes a tricritical point, and the vicinity of it is subject to substantial uncertainty. Five representative points (P1, P2, $\cdots$, P5) on the the phase boundaries are selected, and the corresponding central charges will be discussed below.}
\label{Fig1}
\end{figure}

From the numerical results, we conclude a rich ground-state phase diagram of the spin-1/2 HIAC model with DM interactions, which is shown in Fig. \ref{Fig1}. One can find four different phases: an AFSP, AF, and two distinct Haldane (odd- and even-Haldane) phases, all separated by critical lines. Two magnetic ordered phases can be identified by local order parameters (stripe and N\'{e}el orders, respectively), while the non-magnetic odd- and even-Haldane phases are distinguished by two nonzero string order parameters. In Fig. \ref{Fig1}, besides the QPTs between different magnetic ordered phases and those between ordered and Haldane phases, we also observe a TQPT between the odd- and even-Haldane phases, which can not be explained via symmetry breaking (Landau paradigm). This TQPT can be well captured by the singular behavior of the bipartite entanglement, as well as  even- and odd- string order parameters.

In addition, we are also interested in the critical properties on the phase boundaries. We select several representative points (P1 to P5, see Fig. \ref{Fig1}) and calculate their central charges $c$ by fitting the block entanglement entropies. We find $c=1/2$ at the magnetic order-disorder transition points (P1, P2, and P3); while $c=1$ at the TQPT points (P4 and P5) between two kinds of Haldane phases .

\section{Numerical results and discussions}
\label{section3}
In the following, we study the phase transitions along two selected paths, by fixing $D$=0.7 or 2.0 and tuning Ising couplings $J_I$. In subsection \ref{section3.1}, we study the line $D$=0.7 crossing four different phases (see Fig. \ref{Fig1}), i.e., the AFSP, odd-Haldane, AF, and even-Haldane phases, and focus on the magnetic order-disorder QPTs. In subsection \ref{section3.2}, we focus on the path along $D$=2.0 line, along which the odd- and even-Haldane phases touch each other and a QPT was found between them. Since both the even- and odd-Haldane phases have the same symmetry properties (no symmetry breaking), this exotic phase transition between two phases can be regarded as a TQPT. in subsection \ref{section3.3}, the block entanglement entropies and the fitted central charges of several critical points are evaluated. In our calculations, the Heisenberg coupling $J_{H}=1$ [see Eq. (\ref{Hamiltonian2})] is set as an energy scale, and the retained bond dimension of MPS is set as $\chi =30$. In order to check the reliability of our results, we reinvestigate the Heisenbeg-Ising chain with odd-bond DM interactions in subsection \ref{section3.4}, and find that the disordered phase corresponds to the odd-Haldane phase. Next we study the effect of moderate strengths of even-bond DM interactions, and the evolution of the even-Haldane phase are shown pictorially in subsection \ref{section3.5}.

\subsection{The QPTs along the $D$=0.7 line}
\label{section3.1}

\begin{figure}
\begin{center}
\includegraphics[width=0.7\linewidth]{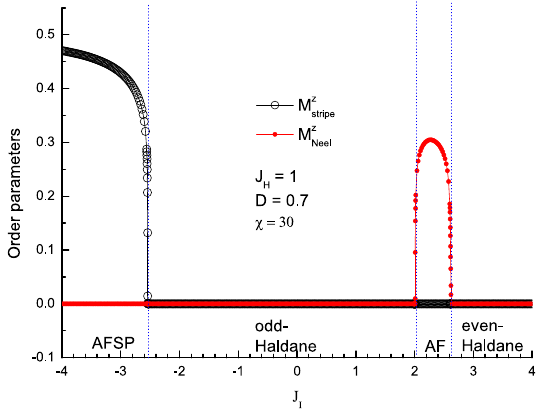}
\end{center}
\caption{(Color online) The stripe order parameter
$M^{z}_{stripe}$ = $\frac{1}{N} |\sum_{i=1}^{N/2} (-1)^{i}(M^z_{2i}+M^z_{2i+1})|$ and the N\'{e}el order parameter $M^{z}_{Neel}$=$\frac{1}{N} | \sum^{N}_{i=1} (-1)^{i} M^z_{i} |$ versus Ising coupling $J_{I}$ are shown (along the $D$=0.7 line).} \label{Fig2}
\end{figure}

In this subsection, we investigate the QPTs along the line $D$=0.7. The local magnetizations on site $i$, $M^{\sigma}_{i}= \langle S^{\sigma}_{i}\rangle$ ($\sigma$=\{$x$, $z$\}) are evaluated. We find that the transverse magnetization $M^{x}_{i}$ vanishes completely along the line. However, local magnetizations $M^{z}_{i}$ are nonzero in two regions: In the region of $J_{I}<-2.54$, the spin configuration is $\cdots+--+\cdots$ or $\cdots-++-\cdots$ ( + and - denote spin up and down in terms of $M^{z}_{i}$, respectively), constitutes a period-four stripe-like AF ordered phase (dubbed as AFSP); another symmetry broken phase is between $2.0<J_{I}<2.6$, where we observe an conventional AF N\'{e}el order, i.e., $\cdots+-+-\cdots$ or  $\cdots-+-+\cdots$. It is convenient to define the stripe order parameter $M^{z}_{stripe}$ = $\frac{1}{N} |\sum_{i=1}^{N/2} (-1)^{i}(M^z_{2i}+M^z_{2i+1})|$ and the N\'{e}el order parameter $M^{z}_{Neel}$=$\frac{1}{N} | \sum^{N}_{i=1} (-1)^{i} M^z_{i} |$ to distinguish these two magnetic ordered phases. $M^z_{Neel}$ and $M^z_{stripe}$ are evaluated for various $J_{I}$ and shown in Fig. \ref{Fig2}. As expected, $M^{z}_{stripe}$ and $M^{z}_{Neel}$ can be used to distinguish the AFSP and AF phase; however, they are not able to tell the differences between the rest two nonmagnetic regions ($-2.54<J_I<2$ and $J_I>2.6$), where local magnetic moments vanish and no symmetry breaking is observed. Interestingly, by introducing some nonlocal order parameters, we see that these two regions ($-2.54<J_I<2$ and $J_I>2.6$) have different types of non-local string order, suggesting the existence of two different SPT orders. We will discuss these two distinct phases, namely the odd- and even-Haldane phases, and the phase transition between them later in subsection \ref{section3.2}.

\begin{figure}
\begin{center}
\includegraphics[width=0.7\linewidth]{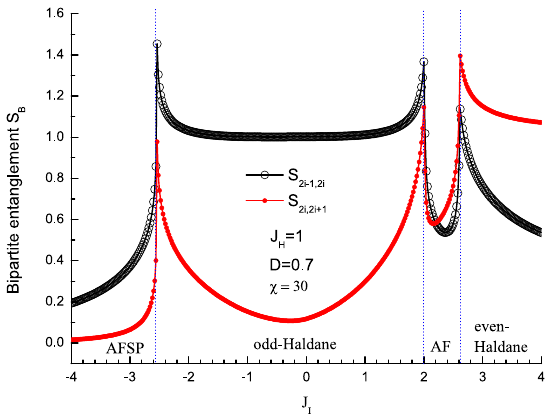}
\end{center}
\caption{(Color online) Bipartite entanglement with $D$=0.7 on odd bond ($S_{2i-1,2i}$) and even bond ($S_{2i,2i+1}$).} \label{Fig3}
\end{figure}

\begin{figure}
\begin{center}
\includegraphics[width=0.7\linewidth]{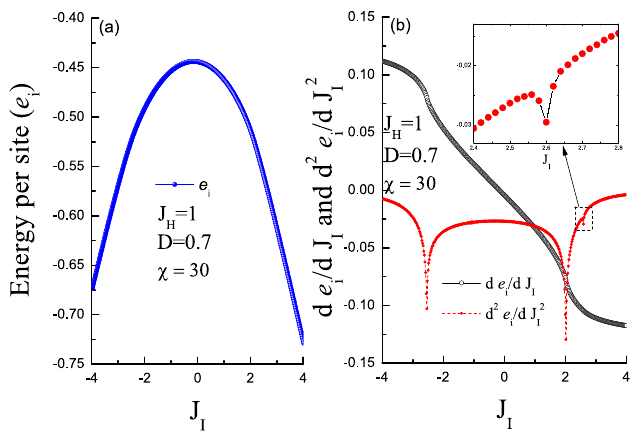}
\end{center}
\caption{(Color online) (a) Ground-state energy per site $e_{i}$ with $D$=$0.7$; (b) The first- and second-order derivative curves of $e_{i}$.} \label{Fig4}
\end{figure}

The entanglement entropy has been shown to be an efficient tool to detect QPTs \cite{Amico,Liu1,You14}. For continuous QPTs, owing to the quantum criticality at the transition point, the bipartite von Neumann entanglement entropy
\begin{equation}
S_{\rm{B}} = - \rm{Tr} (\rho \log_2 \rho),
\label{Definition Of Entanglement}
\end{equation}
diverges at the critical point. In the MPS framework, $S_{\rm{B}}$ can be evaluated by $S_{\rm{B}} = - \sum_{i=1}^{\chi} \Lambda_{i}^{2} {\rm \log_2}  \Lambda_{i}^{2}$, where $\Lambda_i$'s are (normalized) singular values obtained through bond singular value decomposition, i.e., square root of the eigenvalues of the half-infinite-chain reduced density matrix. In Fig. \ref{Fig3}, we show the results of the bipartite entanglement entropies $S_{2i-1,2i}$ on the odd bond and $S_{2i,2i+1}$ on the even bond. $S_{\rm{B}}$ shows three sharp peaks, indicating three continuous QPT points. Besides, we have also calculated the ground-state energy per site and and its derivatives (versus $J_I$), and show them in Fig. \ref{Fig4}. Although the ground-state energy per site $e_{i}$ and its first-order derivative ($d e_{i}/d J_{I}$) behave continuously [Fig.~\ref{Fig4} (a) and (b), respectively], the second-order derivative ($d^{2} e_{i}/d J_{I}^{2}$) exhibits divergent peaks at three critical points [Fig.~\ref{Fig4} (b)]. The singularities in $d^{2} e_{i}/d J_{I}^{2}$ confirm the conclusion that these three magnetic order-disorder QPTs are of second order \cite{Liu}.

\subsection{The $D$=2 line and the topological phase transition}
\label{section3.2}

\begin{figure}
\begin{center}
\includegraphics[width=0.7\linewidth]{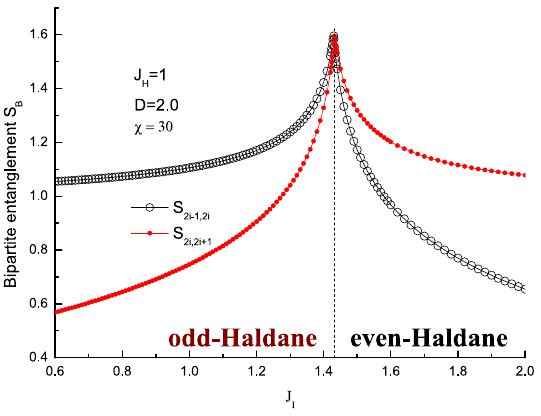}
\end{center}
\caption{(Color online) Bipartite entanglement on odd bond ($S_{2i-1,2i}$) and even bond ($S_{2i,2i+1}$) with $D$=2.0.} \label{Fig5}
\end{figure}

Next, we explore the line $D$=2 in Fig. \ref{Fig1}, along which the bipartite entanglement entropy is calculated (see Fig. \ref{Fig5}). We will focus on the QPT occurred at $J^{c}_{I}\simeq1.43$ (P4 point of Fig. \ref{Fig1}), between two magnetic disordered phases, i.e., the odd- and even-Haldane phases. The sharp peak of the entanglement entropy curve suggests that this QPT at P4 is also of second order.

We firstly show the results of the nonlocal string order parameters, which is introduced by den Nijs and Rommels \cite{Nijs}, and are used to characterize the dilute AF order in the Haldane phase of the spin-1 Heisenberg chain. The presence of a nonzero string order parameter can be explained via a hidden $Z_2 \times Z_2$ symmetry breaking \cite{Tasaki,Kennedy}, and be related to a symmetry-protected topological order. For the present model, in order to distinguish the odd- and even-Haldane phases, we need to define two different string orders (odd- and even-string orders), which are defined as \cite{Wang, Liu}
\begin{eqnarray}
&O^{\alpha,odd}_{s} (L) &=\langle \sigma^{\alpha}_{2i-1} \sigma^{\alpha}_{2i} \cdots  \sigma^{\alpha}_{2j-1} \sigma^{\alpha}_{2j} \rangle, \\
&O^{\alpha,even}_{s} (L)&= \langle \sigma^{\alpha}_{2i} \sigma^{\alpha}_{2i+1} \cdots  \sigma^{\alpha}_{2j} \sigma^{\alpha}_{2j+1} \rangle,
\label{stringorder}
\end{eqnarray}
in which $L=|j-i|$, and $\sigma^{\alpha}$'s ($\alpha$=$x$, $y$ and $z$) denote spin-1/2 pauli matrices. The $O^{\alpha,odd}_{s}$ is calculated from an odd site ($2i-1$) to an even site ($2j$), but the $O^{\alpha,even}_{s}$ starts from an even site ($2i$) and ends at an odd site ($2j+1$), i.e., the lattice distance is $L$=2$(j-i+1)$ in both cases.

\begin{figure}
\begin{center}
\includegraphics[width=0.7\linewidth]{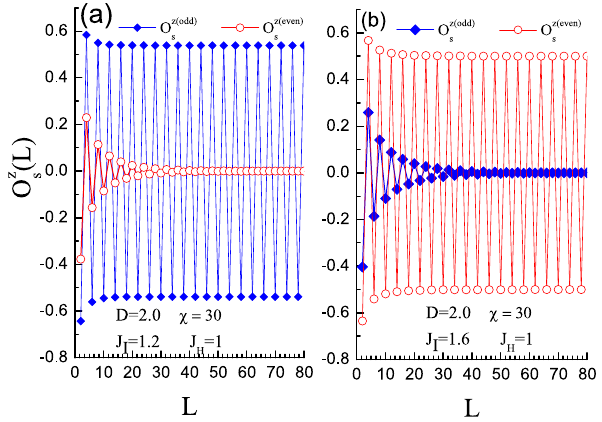}
\end{center}
\caption{(Color online) The string order parameters $O^{z(odd)}_{s} (L)=\langle \sigma^{z}_{2i-1}
\sigma^{z}_{2i} \cdots \sigma^{z}_{2j-1} \sigma^{z}_{2j} \rangle$ and $O^{z(even)}_{s} (L)=\langle \sigma^{z}_{2i}
\sigma^{z}_{2i+1} \cdots \sigma^{z}_{2j} \sigma^{z}_{2j+1} \rangle$ with: (a) $J_{H}$=1, $J_{I}$=1.2, and $D$=2 (a point in the odd-Haldane phase); (b) $J_{H}$=1, $J_{I}$=1.6, and $D$=2 (a point in the even-Haldane phase).  $L = |j-i|$ is the length of the string, and $\sigma$ operator denotes a spin-1/2 Pauli matrix.} \label{Fig6}
\end{figure}

\begin{figure}
\begin{center}
\includegraphics[width=0.7\linewidth]{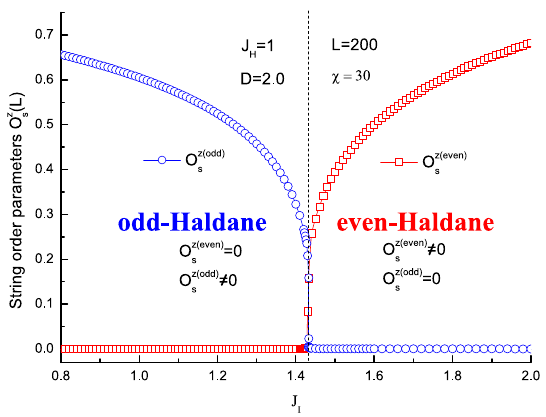}
\end{center}
\caption{(Color online) Absolute values of $O^{z(odd)}_{s} (L)$ and $O^{z(even)}_{s} (L)$ (at $L$=200), which serve as order parameters characterizing the odd- and even-Haldane phases.} \label{Fig7}
\end{figure}

\begin{figure}
\begin{center}
\includegraphics[width=0.7\linewidth]{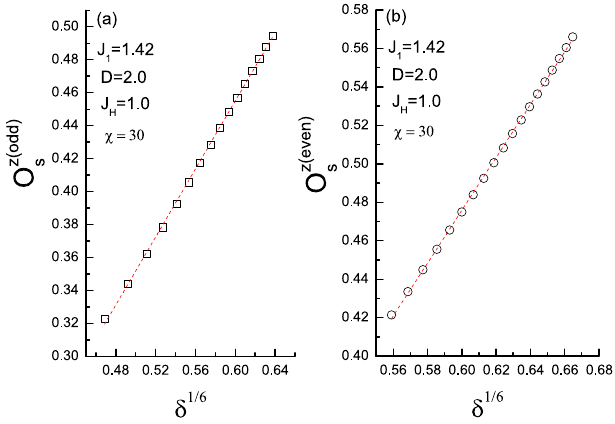}
\end{center}
\caption{(Color online) String order parameters (a) $O^{z(odd)}_{s}$ and (b)
$O^{z(even)}_{s}$
  as a function of $\delta = J_I-J_I^c $.
 } \label{Fig8}
\end{figure}

In Fig.~\ref{Fig6}, we show the results of both the odd-string $O^{z,odd}_{s}$ and even-string $O^{z,even}_{s}$ order parameters at two representative points ($J_{H}$=1, $J_{I}$=1.2, and $D$=2) and ($J_{H}$=1, $J_{I}$=1.6, and $D$=2), which locate at distinct nonmagnetic phases (Fig. \ref{Fig1}). From Fig. \ref{Fig6}, we find that $O^{z,odd}_{s} (L)$ and $O^{z,even}_{s} (L)$ show different behaviors at these two points: The order parameter $O^{z,even}_{s}$ ($O^{z,odd}_{s}$ ) in the odd-Haldane (even-Haldane) phase decay quickly to zero within the lattice distance $L$=50 (see Fig. \ref{Fig6}) ; On the other hand, $O^{z,odd}_{s}$ ($O^{z,even}_{s}$ ) in the odd-Haldane (even-Haldane) phase oscillates, whose absolute values converge to nonzero values in the large $L$ limit (Fig. \ref{Fig6}). Collecting the converged (absolute) values of $O^{z, odd}_{s}$ and $O^{z, even}_{s}$ and show them in Fig. \ref{Fig7}, one can clearly see that $O^{z, odd}_{s}$ and $O^{z, even}_{s}$ serve as well-defined order parameters of odd- and even-Haldane phases, respectively. In addition, in the vicinity of the critical point, we find the (converged) string orders obey a power-law scaling versus parameter $\delta \equiv \vert J_{I}-J^{c}_{I}\vert$:
\begin{equation}
O^{z}_{s} \sim    \delta ^{2\beta},
\end{equation}
where $\beta=1/12$ [see Fig. \ref{Fig8} (a) and (b)].

\begin{figure}
\begin{center}
\includegraphics[width=0.7\linewidth]{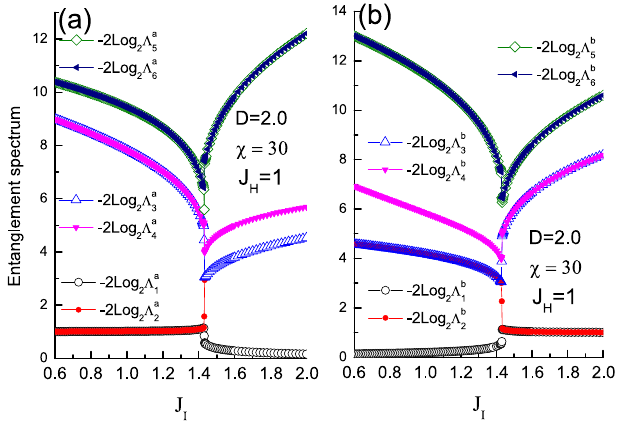}
\end{center}
\caption{(Color online)  Entanglement spectra on (a) odd bond (-2 Log$_{2}\Lambda^{a}_{i}$) with $D$=$2.0$ and (b) on even bond (-2 Log$_{2}\Lambda^{b}_{i}$) along the line $D$=$2.0$.} \label{Fig9}
\end{figure}

Subsequently, we analyze the entanglement spectrum (ES) \cite{Li-2008}, which is defined as the -Log$_2(\rho)$, where $\rho$ is the reduced density matrix of half-infinite chain. In the MPS framework, the ES can be evaluated through -2 Log$_{2}\Lambda^{a}_{i}$ (-2 Log$_{2}\Lambda^{b}_{i}$), corresponding to the ES by cutting an odd or even bond. The $\Lambda^{a}$ and $\Lambda^{a}$ denote diagonal matrices on odd and even bonds. The ES results of odd (even) bonds are shown in Fig. \ref{Fig9} (a) and (b), respectively. Different spectrum structures in the odd- and even-Haldane phases are observed. In the odd-Haldane phase, doubly degenerate ES exists on odd bonds (see Fig. \ref{Fig9} (a)). However, in the even-Haldane phase, the ES becomes doubly degenerated on even bonds, instead. It means that the odd- and even-Haldane phases can be characterized by doubly degenerate ES on odd and even bonds, respectively. The degeneracy of ES is a reminiscence of that in spin-1/2 one-dimensional quantum compass models, where there exists a disordered phases with nonzero string orders \cite{Motamedifar,Liu}. The double degeneracy of the ES reflects that the geometric bond space supports a projective representation of global symmetries, including the bond centered inversion symmetry \cite{Pollmann-2010}, the rotational SO(3) symmetry \cite{Li-2013}, etc.

\begin{figure}
\begin{center}
\includegraphics[width=0.7\linewidth]{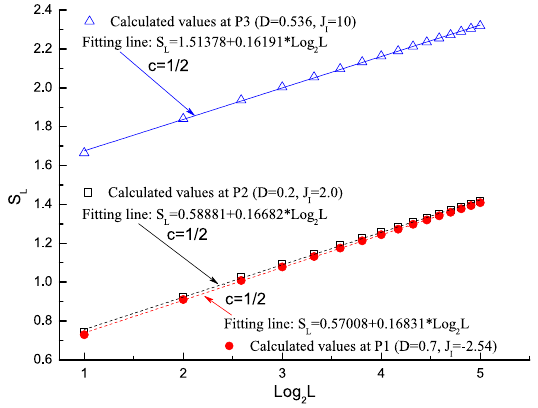}
\end{center}
\caption{(Color online) Block entanglements ($S_L$) with $J_H = 1$ at P1 ($D$=0.7, $J_{I}$=-2.54), P2($D$=0.2, $J_{I}$=2.0), and P3 ($D$=0.536, $J_{I}$=10)  (see P1, P2, and P3 in Fig. \ref{Fig1}). The block entanglements of these points are fitted well by function $k+\frac{c}{3}$Log$_{\rm 2}$L, and their central charges (c) are determined to be about $1/2$.} \label{Fig10}
\end{figure}

\begin{figure}
\begin{center}
\includegraphics[width=0.7\linewidth]{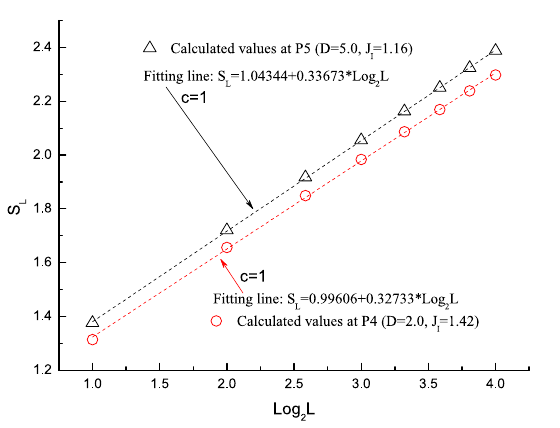}
\end{center}
\caption{(Color online) Block entanglements ($S_L$) with $J_H = 1$ at P4 ($D$=2.0, $J_{I}$=1.42) and P5($D$=5.0, $J_{I}$=1.16). The block entanglement entropies of both points are fitted well by $k+\frac{c}{3}$Log$_{\rm 2}$L, and their central charges are determined to be $c\simeq1$.} \label{Fig11}
\end{figure}

\subsection{BLOCK ENTANGLEMENT ENTROPY AND CENTRAL CHARGE}
\label{section3.3}
From the phase diagram shown in Fig. \ref{Fig1}, one can find that there exist four different phases, and all the QPTs on the phase boundaries separating them are found to be critical. In order to study their criticalities, in Figs. \ref{Fig10},\ref{Fig11} we show the block entanglement entropy $S_{L}$, which measures the entanglement between a block of $L$ spins and the remaining (infinite) environment. The block entanglements of gapped phases are found to be saturated when $L$ is large enough, well satisfying the so-called area law \cite{Eisert}. However, for the critical points, the block entanglement $S_{L}$ has a logarithmic correction to area law, as derived in Ref. \cite{Holzhey}, $S_{L}$ increases with $L$ as
\begin{equation}
S_{L}=k+\frac{c+\bar{c}}{6}{\rm \log_{2}L},
\label{block entanglement}
\end{equation}
where $k$ denotes the intercept and $c$ ($\bar{c}$) is the holomorphic (antiholomorphic) central charge in the conformal field theory (CFT). The central charges $c$ and $\bar{c}$ are important characterizations of CFT, and can be used to classify the universality class of QPTs \cite{Calabrese}.

Five representative points labeled as $P1$, $P2$, $\cdots$, $P5$ (see Fig. \ref{Fig1}) are selected from the phase boundaries as examples. Among them, $P1$, $P2$, and $P3$ are three points on the phase boundaries between two Haldane phases and the AFSP and AF phases, and $P4$, and $P5$ are critical points between the odd- and even-Haldane phases. The block entanglement entropies of $P1$, $P2$, and $P3$ are plotted in Fig.~\ref{Fig10},
where logarithmic divergent behaviors are observed at these three points, and the central charges are determined to be $c \simeq$ 1/2. It means the QPTs from two Haldane phases to the AFSP and AF phases belong to the same universality class, i.e., the Ising universality class, and they can be described by a free fermionic field theory \cite{Vidal2003}, with central charges $c$=$\bar{c} \simeq 1/2$. On the other hand, logarithmic divergent behaviors are also observed at $P_4$ and $P_5$ (see Fig.~\ref{Fig11}), but with central charges determined as $c \simeq1$, showing that the topological phase transition between the odd- and even-Haldane phases belongs to the Gaussian universality class, and may be described by a free bosonic field theory.

\subsection{Revisit of the HIAC with DM interactions on odd bonds only}
\label{section3.4}
\begin{figure}
\begin{center}
\includegraphics[width=0.7\linewidth]{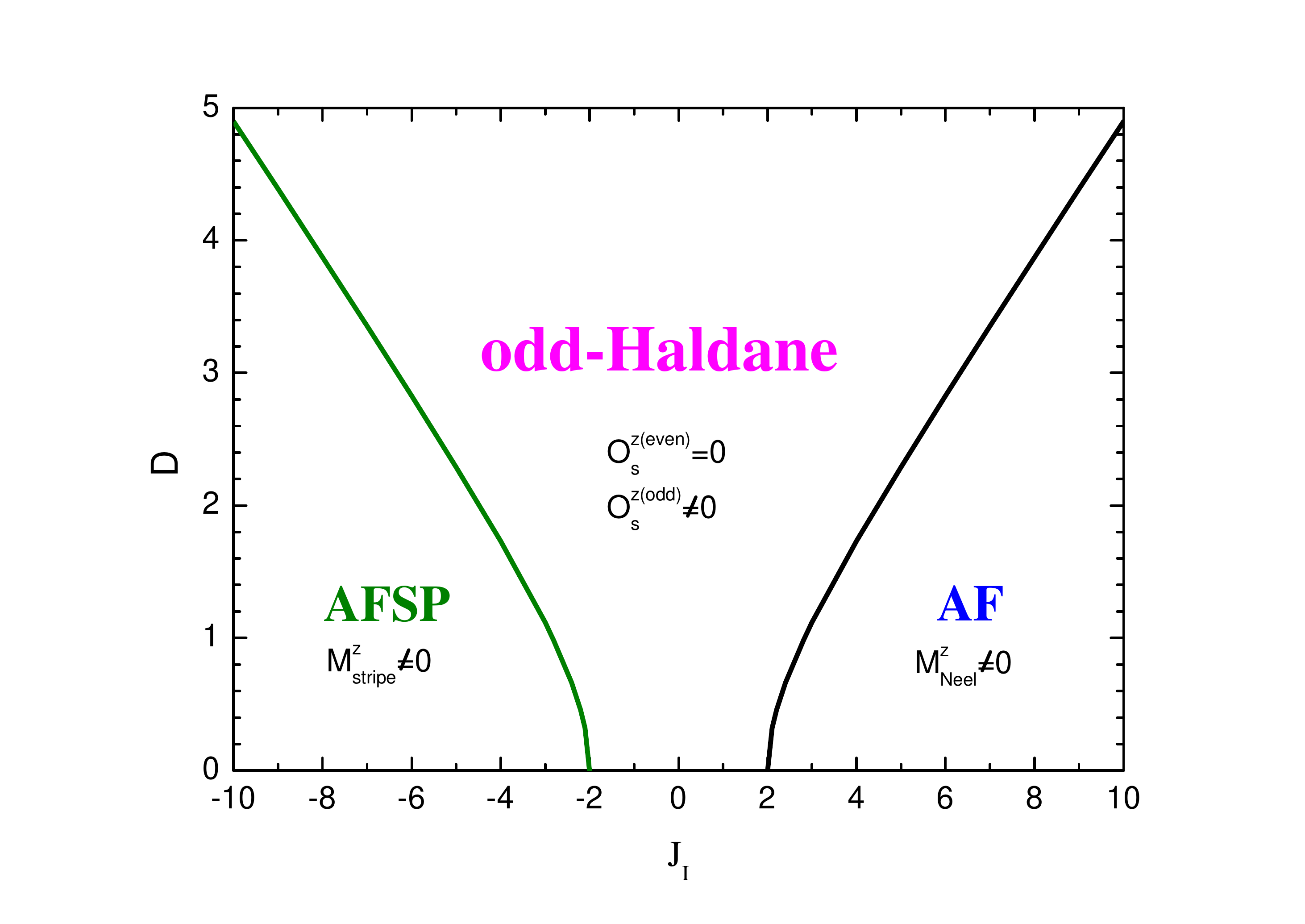}
\end{center}
\caption{(Color online) Phase diagram of a spin-1/2 HIAC with DM interactions on odd bonds [Eq. (\ref{Hamiltonian3})]. It includes three different phases: an AFSP, an AF phase, and an odd-Haldane phase. Two second-order critical lines locate exactly at $J^{c}_{I}$=$\pm 2\sqrt{D^{2} + J^{2}_{H}}$ ($J_{H}$ is set to be 1 in our calculations).} \label{Fig12}
\end{figure}

In this subsection, we switch off the DM interaction on even bonds in Hamiltonian (\ref{Hamiltonian2}), and thus reduce the model as\begin{eqnarray}
 \hat{H} &=& \sum^{N/2}_{i} [J_{H} {\bold{S}}_{2i-1} \cdot {\bold{S}}_{2i}+ D (S^{x}_{2i-1} S^{y}_{2i}-S^{y}_{2i-1} S^{x}_{2i}) \nonumber \\
         &+& J_{I} S^{z}_{2i} S^{z}_{2i+1}].
\label{Hamiltonian3}
\end{eqnarray}
As shown in Refs. \cite{Lieb,Yao,Strecka}, this model is exactly solvable. Recently Derzhko \emph{et al.}calculated the ground-state compressibility of a deformable spin-1/2 Heisenberg-Ising chain with DM interaction on odd bonds only \cite{Derzhko}. In Fig.~\ref{Fig12}, we revisit its ground-state phase diagram through numerical simulations. Two phase boundaries detected by the sharp peaks of bipartite entanglement locate at $J^{c}_{I}$=$\pm 2\sqrt{D^{2} + J^{2}_{H}}$, which perfectly agrees with analytical result \cite{Strecka}. As $|J_{I}| < 2\sqrt{D^{2} + J^{2}_{H}}$, it is a disordered phases with nonzero $O^{z,odd}_{s}$ but with vanishing $O^{z,even}_{s}$. Furthermore, the doubly degenerate ES is observed on the odd bonds. These two facts, nonzero $O^{z,odd}_{s}$ and doubly degenerate ES, indicate that the disordered phase is an odd-Haldane phase. In the region $J_{I} < -2\sqrt{D^{2} + J^{2}_{H}}$, an AFSP is observed; while an AF phase appears when $J_{I} > 2\sqrt{D^{2} + J^{2}_{H}}$. From the Hamiltonian (\ref{Hamiltonian3}), we realize that the case with positive $J_{I}$ can also be connected to that with negative $J_{I}$ by a unitary
transformation. More specifically, the Hamiltonian with $-J_{I}$ and that with $J_{I}$ can be mutually transformed by the $\pi$-rotation of the spins at sites $4i-3$ and $4i-2$ (or equivalently at $4i-1$ and $4i$) around the $x$- or $y$-axis. For instance, if all the spins at sites $4i-3$ and $4i-2$ are rotated about $x$-axis by $\pi$, one can get
\begin{eqnarray}
&& S_{4i-3(4i-2)}^{x} \rightarrow S_{4i-3(4i-2)}^{x}, \nonumber \\
&& S_{4i-3(4i-2)}^{y} \rightarrow -S_{4i-3(4i-2)}^{y}, \nonumber \\
&& S_{4i-3(4i-2)}^{z} \rightarrow -S_{4i-3(4i-2)}^{z}.
\end{eqnarray}
After this unitary transformation, the new Hamiltonian is identical to the old one, but with $-J_{I}$. Therefore, when all the spins at sites $4i-3$ and $4i-2$ are rotated over $x$-axis by $\pi$, the spin configuration  $\cdots-+-+-+-+\cdots$ of the AF phase in the region $J_{I} > 2\sqrt{D^{2} + J^{2}_{H}}$ will become an AFSP configuration  $\cdots+--++--+\cdots$  in the region $J_{I} < -2\sqrt{D^{2} + J^{2}_{H}}$, in accordance with our numerical observation.

\subsection{Evolution of the even-Haldane phase}
\label{section3.5}
\begin{figure}
\begin{center}
\includegraphics[width=0.7\linewidth]{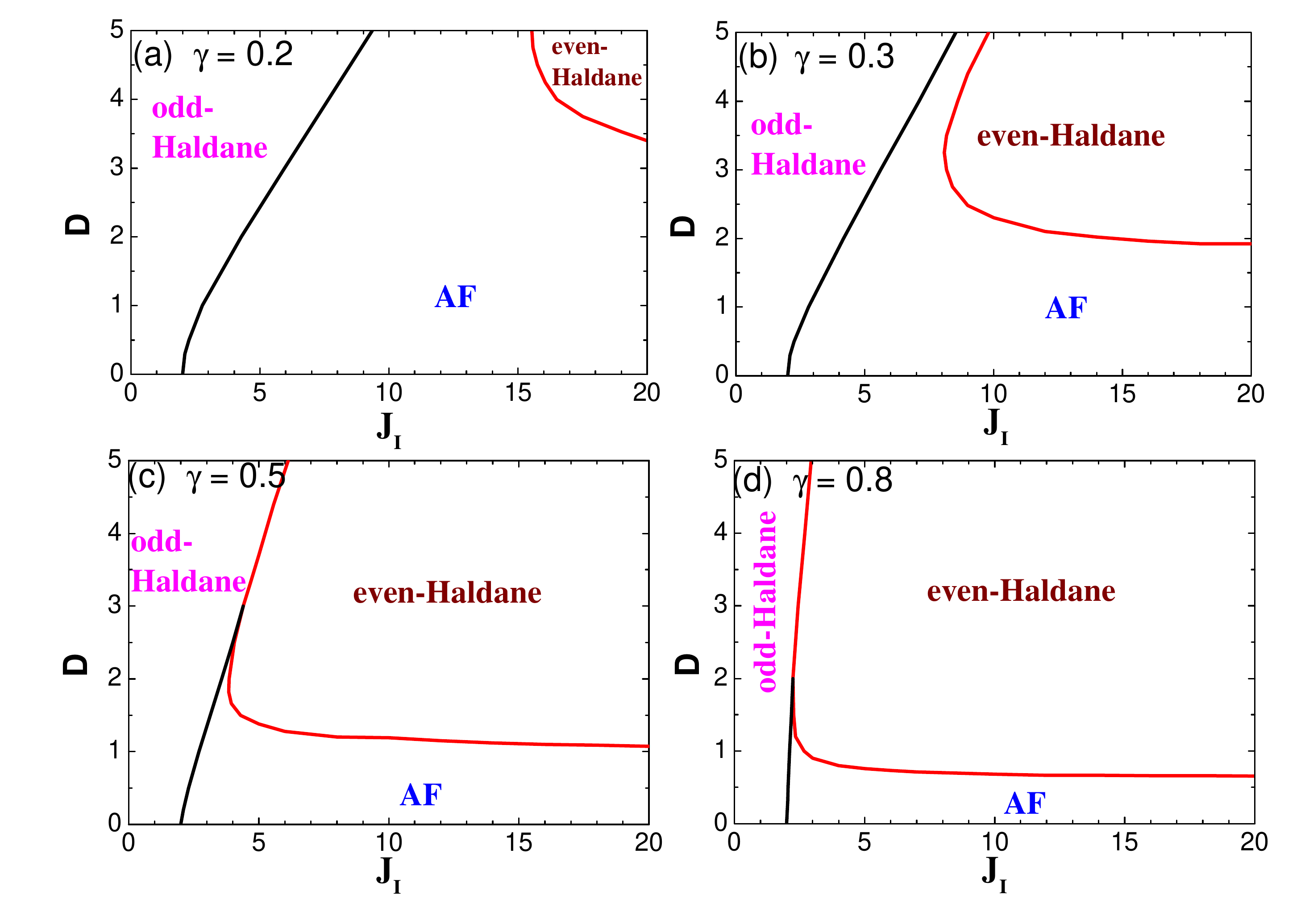}
\end{center}
\caption{(Color online)  Phase diagrams of a spin-1/2 HIAC with DM interactions on odd and even bonds for (a) $\gamma$=0.2; (b) $\gamma$=0.3; (c) $\gamma$=0.5 and (d) $\gamma$=0.8. The $\gamma$ denotes the relevant strength of the DM interactions on even bonds compared with that on odd bonds [see definition in Eq. (\ref{Hamiltonian4})].} \label{Fig13}
\end{figure}

From the phase diagrams in Fig.~\ref{Fig1} and Fig.~\ref{Fig12}, we find that an even-Haldane phase will be induced as the DM interactions on even bonds is switched on. In order to explore the evolution of such an even-Haldane phase, we introduce the following Hamiltonian with alternating DM interactions to interpolate both limiting cases:
\begin{eqnarray}
 \hat{H} &=& \sum^{N/2}_{i} [J_{H} \hat{\bold{S}}_{2i-1} \cdot \hat{\bold{S}}_{2i}+ D (S^{x}_{2i-1} S^{y}_{2i}-S^{y}_{2i-1} S^{x}_{2i}) \nonumber \\
         &+& J_{I} S^{z}_{2i} S^{z}_{2i+1}+ \gamma D (S^{x}_{2i} S^{y}_{2i+1}-S^{y}_{2i} S^{x}_{2i+1})].
\label{Hamiltonian4}
\end{eqnarray}
The $\gamma$ denotes the relevant strength of the DM interactions on even bonds compared with that on odd bonds, and thus the values of interest are restricted within the range $[0,1]$. When $\gamma$=0, it reduces to the Hamiltonian (\ref{Hamiltonian3}) with DM interactions on odd bonds only and turns into Hamiltonian (\ref{Hamiltonian2})
when $\gamma$=1. In addition to the studied cases of $\gamma$=0 and 1, we also select $\gamma$=0.2, 0.3, 0.5 and 0.8 to demonstrate the evolution
of the even-Haldane phase. The corresponding phase diagrams by tuning $D$ and $J_{I}$ are provided in Figs.~\ref{Fig13} (a-d), respectively. Since
the effect of the even-bond DM interactions on the AFSP is negligible (see Fig.~\ref{Fig1} and Fig.~\ref{Fig12}), we just focus on the
region $J_{I}>$0. We find that, an even-Haldane phase can be induced as long as the DM interactions on even bonds are taken into account. When $\gamma$ is small, the even-Haldane phase locates in the region with very large $D$ and $J_{I}$ (figures are not shown here). When $\gamma$ increases, the even-Haldane phase expands quickly while the AF and the odd-Haldane phases shrink. As $\gamma$ is large enough, this even-Haldane phase comes into the region with small $D$ and $J_{I}$ [see Fig.~\ref{Fig13} (a-d)]. Because the even-Haldane phase, the odd-Haldane phase and AF phase are adjacent to each other, a tricritical point exists between them. As $\gamma$ increases, such a tricritical point moves towards the region with
small $D$ and $J_{I}$, and appears in the phase diagram with small $D$ and $J_{I}$ consequently (see Fig.~\ref{Fig13} (c) and (d)). It is worth noting that the exact position of such a tricritical point is difficult to be determined precisely.

\section{DISCUSSIONS AND CONCLUSION}
\label{sec4}

The ground-state phase diagram and the QPTs in the spin-1/2 HIAC with uniform DM interactions have been investigated by the MPS method. By calculating the odd- and even-string order parameters, two kinds of Haldane phases, i.e., odd- and even-Haldane phases, have been identified. Furthermore, doubly degenerate entanglement spectra on odd and even bonds are observed in odd- and even-Haldane phases, respectively. A rich phase diagram including four different phases, i.e., an AFSP, an AF phase, odd- and even-Haldane phases, have been obtained. The TQPT between the odd- and even-Haldane phases is with central charge $c$=1 (a Gaussian type phase transition) and with critical exponent $\beta=1/12$. The central charges on the other phase boundaries are $c$=1/2, therefore the QPTs from nonmagnetic (two Haldane) phases to the magnetic ordered (AFSP and AF) phases belong to the Ising universality class.


In addition, the bipartite entanglement entropy has been shown as a very powerful tool for capturing QPTs (including TQPTs). Two kinds of nonlocal string orders and the doubly degenerate ES can be used to distinguish the odd- and even-Haldane phases. The nonzero string order parameters imply the breaking of the hidden topological symmetry in the odd- and even-Haldane phases. It is worth noticing that the existence of the odd-Haldane phase has been already proved for the Heisenberg-Ising chain without Dzyaloshinskii-Moriya interaction \cite{Liugh}. We would like to remark that the tricritical point (see the filled square $\blacksquare$ in Fig. \ref{Fig1}) is rather difficult to perform an accurate calculation, posing a challenging problem in numerical simulations. In order to explore the evolution of the even-Haldane phase, we introduced a Hamiltonian with a tunable DM interactions on even bonds. Our results show that the even-Haldane phase can be induced so long as the DM interactions on even bonds are taken into account. The DM interactions on even bonds plays a key role in the formation of the even-Haldane phase.

Above we only considered the case with positive DM interaction ($D>0$). In fact, we realize that the case with negative $D$ is connected to that with positive $D$ via a transformation: one can rotate all the spins by $\pi$ angles about the x-axis, \textit{i.e.}, $S^{x} \rightarrow S^{x}$, $S^{y} \rightarrow -S^{y}$ and $S^{z} \rightarrow -S^{z}$. This rotation leaves the Hamiltonian intact, but adds a minus sign before the DM term. Therefore, it is sufficient to consider only the case of positive $D$.

{\bf ACKNOWLEDGMENT}
This work is supported by the Chinese National Science Foundation under Grant No.~11347008 and No.~11474211. It is also partially supported by the National Basic Research Program of China under Grant No.~2012CB932900. W.-L.Y. acknowledges support by the Natural Science Foundation of Jiangsu Province of China under Grant No. BK20141190.

\end{document}